\documentclass[secnumarabic,prb,reprint,groupedaddress]{revtex4-1}
\usepackage{verbatim}
\usepackage[english]{babel}
\usepackage{graphicx}
\usepackage{color}
\usepackage{amsmath}

\begin{document}

\title{Experimental demonstration and analysis of random field effects in ferromagnet/antiferromagnet bilayers}

\author{Guanxiong Chen$^1$}
\author{Dylan Collette$^1$}
\author{Sergei Urazhdin$^1$}

\affiliation{$^1$Department of Physics, Emory University, Atlanta, GA, USA.}

\begin{abstract}
More than 30 years ago, Malozemoff (Phys. Rev. B {\bf 35}, 3679 (1987)) hypothesized that exchange interaction at the interface between a ferromagnet (F) and an antiferromagnet (AF) can act as an effective random field, which can profoundly affect the magnetic properties of the system. However,  until now this hypothesis has not been directly experimentally tested. We utilize magnetoelectronic measurements to analyze the effective exchange fields at Permalloy/CoO interface. Our results cannot be explained in terms of quasi-uniform effective exchange fields, but are in agreement with the random-field hypothesis of Malozemoff. The presented approach opens a new route for the quantitative analysis of effective exchange fields and anisotropies in magnetic heterostructures for memory, sensing and computing applications.
\end{abstract}

\maketitle

\section{Introduction}

The exploration of ferromagnet/antiferromagnet (F/AF) heterostructures started over 60 years ago with the discovery, by Meiklejohn and Bean, of exchange bias (EB) effect - asymmetry of the ferromagnetic hysteresis loop that emerges below a certain blocking temperature $T_B$~\cite{Meiklejohn56}. EB can be utilized for ``pinning" the magnetization of Fs, which has found extensive applications in magnetoelectronic sensors and memory devices~\cite{heck2013magnetic,Akerman508,hu2011high,hu2011high,chappert2010emergence}. A recent resurgence of interest in the fundamental properties of F/AF heterostructures has been motivated by the emergence of AF spintronics - a research field that aims to take advantage of the vanishing magnetization of AFs, their high characteristic dynamical frequencies, and weak coupling to external fields to develop efficient, fast, and stable magnetic nanodevices~\cite{Jungwirth2016}. While some of the implementations of such AF-based devices rely on standalone AFs~\cite{SinovaJairo2012Nmat,Nishitani2010,PhysRevB.95.035422}, many others utilize auxiliary Fs, usually in F/AF heterostructures, to generate spin currents for nanodevice operation, detect the state of AFs, and/or directly control this state via exchange interaction~\cite{PhysRevLett.92.247201,PhysRevB.92.165424,PhysRevLett.113.196602,Fukami2016,PhysRevLett.118.067202,PhysRevB.94.014427,Khymyn2017}. 

Extensive studies of F/AF heterostructures have revealed complex behaviors that sensitively depend on a variety of experimental and material parameters, which could not be explained by na\"ive models assuming perfectly magnetically ordered materials and interfaces~\cite{Nogues1999}. This has lead to the realization that inhomogeneous magnetization states are likely formed in AF and/or F to minimize the exchange energy at the F/AF interfaces. Several  models have been developed to account for this possibility. For instance, some of the observed magnetic properties were attributed to the magnetic domain walls formed in AF to reduce the interfacial exchange energy~\cite{Mauri1987,PhysRevB.66.014430}. It was also proposed that spin glass-like magnetically disordered states can be formed near the F/AF interface~\cite{Schlenker1986,Yamada2007,ding2013interfacial}. 

Even atomic-scale imperfections can reverse the exchange interaction across the F/AF interface, which led Malozemoff~\cite{malozemoff1987random} to suggest that the effects of this interaction can be approximated by an uncorrelated random effective field acting on AF at its interface with F. Analysis based on the extension of the Imry-Ma argument~\cite{PhysRevLett.35.1399} suggested that as a result, AF breaks up into domains. This model predicted EB magnitude qualitatively consistent with the experimental observations. Extending this analysis to ultrathin AF films, Malozemoff also predicted a crossover to the ``Heisenberg domain state" (HDS), wherein AF magnetic domains shrink to sizes below the AF domain wall width~\cite{PhysRevB.37.7673}. The magnetization of AF is then envisioned to become twisted everywhere, and the long-range magnetic ordering of AF is lost. 

The implications of these predictions for the fundamental properties of F/AF heterostructures have so far received relatively little attention~\cite{Rezende,jimenez2009emergence}. Recent time-domain measurements of magnetization states in F/AF bilayers utilizing several common AF materials have revealed universal power law aging~\cite{urazhdin2015cooperative,PhysRevB.94.024422,Ma2018}. Aging was observed only for AF films with thickness below a certain material-dependent value. Thus, aging was attributed to the emergence of a HDS. Based on the analysis of the dependence of aging on the magnetic history and temperature, it was conjectured that in terms of the dynamical properties, the HDS is a correlated spin glass~\cite{Ma2018}. This conjecture was supported by measurements of ac susceptibility, which demonstrated that the temperature dependence of the dynamical response is consistent with the glass transition at the EB blocking temperature $T_B$~\cite{urazhdin2019JMMM}. In particular, the magnetization exhibited viscous dynamics above $T_B$ and elastic dynamics below $T_B$, with viscosity varying by several orders of magnitude close to this temperature. These recent results highlighted the potential significance of the random-field effects proposed by Malozemoff, but have not directly demonstrated the existence of random effective exchange fields at F/AF interfaces.

If the effects of exchange interaction across the F/AF interface can be described by an effective random field exerted on AF, then its reciprocal effects on F can be similarly described by an effective random field. Indeed, the Heisenberg exchange interaction preserves rotational symmetry, and therefore the local exchange torques exerted across F/AF interface on AF should be opposite to the local torques exerted by AF on F. Theoretical studies have shown that random fields acting on Fs produce an inhomogeneous magnetization state, with the magnitude of deviations from the saturated state related to the external field by certain scaling exponents dependent on the system dimensionality~\cite{Chudnovsky1983,PhysRevB.88.224418,PhysRevLett.112.097201,garanin2015ordered}.

Here, we present experimental characterization and analysis of effective exchange fields in Permalloy(Py)/CoO bilayers, one of the ``classic" F/AF bilayer systems extensively studied in the context of EB. In the next section, we introduce our approach. In Section~\ref{sec:exp}, we present measurements of the effects of the applied field on the magnetization states for different thicknesses $t$ of Py, and show that our results for one of the field directions are inconsistent with the approximation of quasi-uniform effective exchange field produced by CoO. In Section~\ref{sec:2drf}, we present an analytical model for the effects of uncorrelated random field on 2d systems. In Section~\ref{sec:micromag}, we utilize a combination of scaling arguments and micromagnetic simulations to extend our analysis to the thin-film geometry of our experiment. In Section~\ref{sec:analysis}, we use the developed approach to show that our experimental results can be explained in terms of the uncorrelated effective random exchange field exerted on Py at its interface with CoO. We also analyze the temperature dependences of the characteristics extracted from our analysis, and show that they are consistent with prior measurements of similar systems. We conclude with a discussion of the scientific and technological relevance of our results.

\section{Our approach}\label{sec:approach}

Our approach to characterizing the exchange interaction at F/AF interfaces is based on the extension of an idea that the spatial characteristics of effective fields acting on a magnetic system determine the functional form of the magnetization curves, as was demonstrated for the effective  anisotropy field by Tejada et al.~\cite{Tejada1991}. We consider the interactions defining the equilibrium state of the magnetization $\vec{M}(\vec{r})$ of F with thickness $t$ in an F/AF bilayer. We assume that $\vec{M}$ is confined to the film plane (the xy plane) by the demagnetizing effects. We neglect the small magnetocrystalline anisotropy of F=Py, which is negligible compared to the other effects discussed here. We also neglect the effects of dipolar magnetic fields, since the analysis of the data presented below excludes highly inhomogeneous magnetization states where these effects may be significant. This set of approximations is commonly referred to as the standard xy spin model.

The Zeeman interaction of $\vec{M}$ with the in-plane external field $H$ is characterized by the magnetic energy density $\epsilon_Z=-\mu_0\vec{M}\cdot\vec{H}$, where $\mu_0$ is the vacuum permeability. The exchange interaction within F can be described by the Heisenberg energy density $\epsilon_{ex}=\frac{A}{M^2}((\vec{\nabla}\vec{M_x})^2+(\vec{\nabla}\vec{M_y})^2)$, where $A$ is the exchange stiffness. Finally, our analysis must include the effects of exchange interaction at the F/AF interface. At the microscopic level, the Heisenberg exchange energy per atom at the interface is $E_{ex,F/AF}=2J_{F/AF}\left<\vec{s}_F\right>\left<\vec{s}_{AF}\right>$, where $J_{F/AF}$ is the Heisenberg exchange constant characterizing the strength of the interaction across the interface, $\vec{s}_F$ is the spin of the F atom at the interface, and $\vec{s}_{AF}$ is the spin of the nearest-neighbor AF atom. Different local atomic arrangements at the interface introduce a correction factor of order one, which can be absorbed in the definition of $J_{F/AF}$.

The interfacial contribution to the energy density can be interpreted, in the spirit of Weiss's molecular field theory, as an effective field $H_{int}=-2J_{F/AF}\left<\vec{s}_{AF}\right>/g\mu_B$ exerted on the interfacial F spins due to the exchange interaction across the interface. Here, $g=2$ is the g-factor for Py, and  $\mu_B$ is the Bohr magneton. This contribution can be also approximated as an effective spatially-varying field acting on the entire F, if we assume that $t$ is sufficiently small so that the  magnetic configuration of F does not significantly vary through its thickness. This approximation is relaxed in the computational analysis presented later in this paper. For F=Py with fcc crystal structure characterized by the cubic lattice constant $a=0.36$~nm, the area per atom at the (111)-textured interface is $P=a^2/4\sqrt{3}$. The magnetic energy density associated with the exchange interaction across the F/AF interface can then be written as $\epsilon_{ex,F/AF}=-\mu_0\vec{M}(\vec{r})\vec{h}(\vec{r})$, where 

\begin{equation}\label{eq:Heff}
\vec{h}(\vec{r})=\frac{4\sqrt{3}J_{F/AF}\left<\vec{s}_{AF}(\vec{r})\right>}{\mu_0Mta^2}
\end{equation}

is the effective exchange field dependent on the in-plane position $\vec{r}$ but uniform through the thickness of F. The magnetic energy density of F is then

\begin{equation}\label{eq:eF}
\epsilon=-\mu_0\vec{M}(\vec{H}+\vec{h})+\frac{A}{M^2}[(\vec{\nabla}M_x)^2+(\vec{\nabla}M_y)^2].
\end{equation}

Following the notations of Garanin et al.~\cite{PhysRevB.88.224418}, who analyzed the 3d version of a similar xy model, we introduce the angle $\varphi(\vec{r})$ between the magnetization and the field $\vec{H}$, and the angle $\phi(\vec{r})$ between $\vec{h}$ and $\vec{H}$. Minimizing the energy $\int{\epsilon(\vec{r})d^2r}$ with respect to $\varphi(\vec{r})$, we obtain

\begin{equation}\label{eq:angles}
\frac{A}{\mu_0M}\nabla^2\varphi(\vec{r})-Hsin\varphi(\vec{r})=hsin(\varphi(\vec{r})-\phi(\vec{r})).
\end{equation}

This equation can be simplified for sufficiently large $H$, when the magnetization is almost saturated, and $\varphi$ is small. We note that even in this limit, often described as the weak random field approximation~\cite{PhysRevB.88.224418}, the magnitude of $h$ needs not be small compared to $H$. In particular, the component $h\sin\phi$ parallel to $\vec{H}$ can be large (both locally and on average), as is the case for F/AF bilayers, where this component determines the unidirectional and the uniaxial anisotropies associated with exchange bias~\cite{Koon97,Schulthess98}. The component $h_\perp=h\sin\phi$ perpendicular to $\vec{H}$ may also be large if it rapidly varies in space, since its effects on the magnetization are averaged out by the exchange stiffness. Separating the contributions of $h_\parallel$ and $h_\perp$ in Eq.~(\ref{eq:angles}), we obtain

\begin{equation}\label{eq:small_angles}
\frac{A}{\mu_0M}\nabla^2\varphi-\varphi(H+h_\parallel)=-h_\perp.
\end{equation}

We assume that neither the preparation of the magnetic system (such as field-cooling) nor its magnetocrystalline properties favor any particular in-plane direction non-collinear with $\vec{H}$. The symmetry with respect to the direction of $\vec{H}$ implies that the average of $h_\perp$ over a sufficiently large area must vanish, and therefore this quantity must vary in space, changing sign over some characteristic length scale $l_h$. 

\begin{figure}
	\includegraphics[width=\columnwidth]{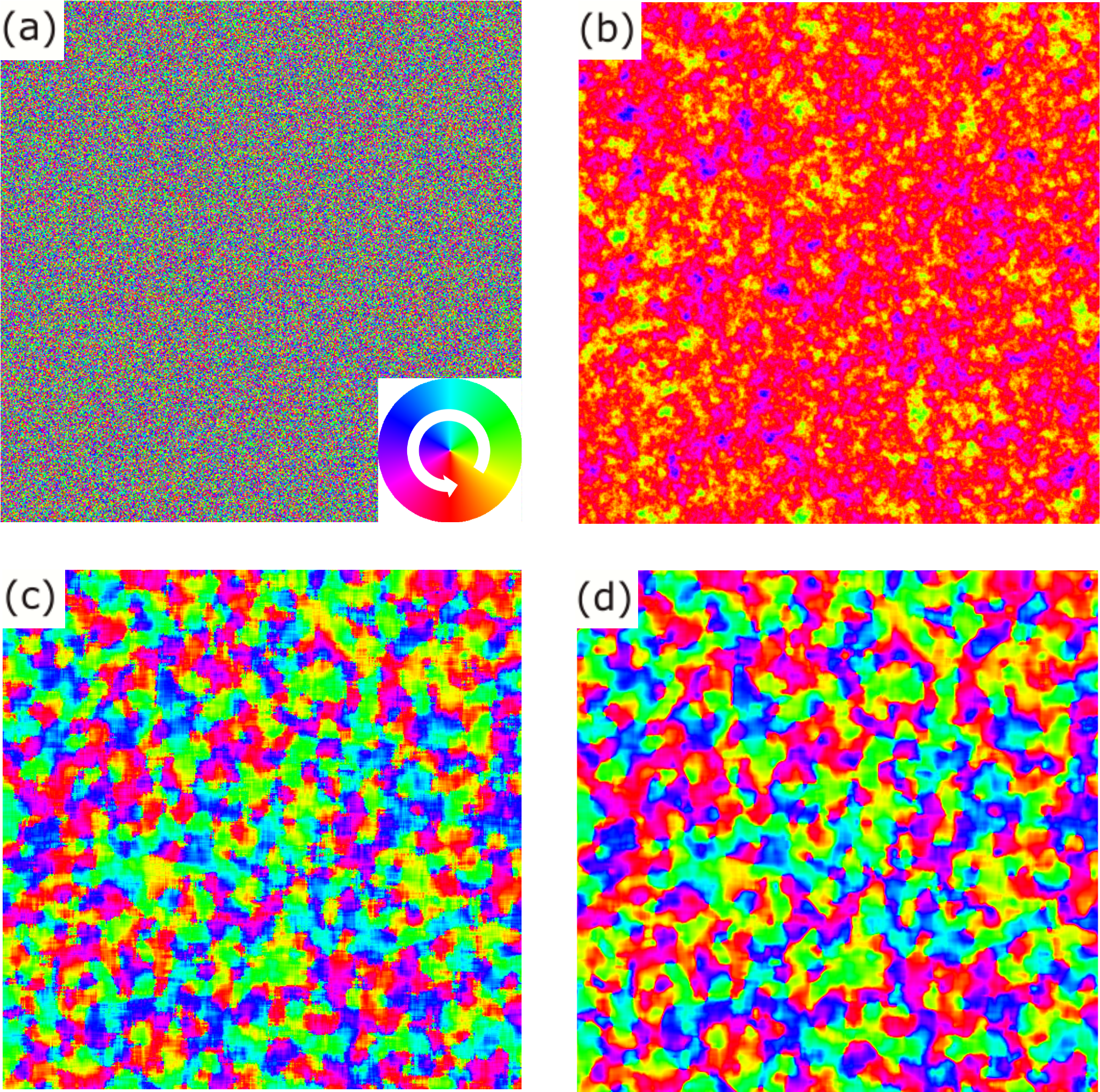}
	\caption{Uncorrelated vs correlated random field effects. (a),(b) Distribution of uncorrelated random field $h=50$~kOe on a 2d mesh of square $2$~nm$\times 2$~nm cells (a) and the resulting magnetization distribution calculated using the $mumax3$ micromagnetic simulation software for a Py(6) film (b), at $H=4$~kOe. For clarity, only a $ 1~\mu m \times 1~\mu m$ region of the $ 2~\mu m \times 2~\mu m$ simulation region is shown. (c),(d) same as (a),(b), for random field with the correlation length $l_h=18$~nm.}\label{fig:randcorr} 
\end{figure}

Malozemoff's uncorrelated random-field approximation is based on the assumption that effective field varies randomly on the atomic lengthscale, i.e. $l_h\sim a$. While the effective field itself is uncorrelated, the exchange stiffness of the ferromagnet defines the magnetic correlation length  $l_M=\sqrt{A/\mu_0M(H+\left<h_\parallel\right>)}$. This is illustrated in Figs.~\ref{fig:randcorr}(a),(b) by the micromagnetic simulations for a Py(6) film subjected to an uncorrelated random field $h=50$~kOe. Here, the number in parenthesis is the thickness in nanometers. The statistical properties of the magnetization state in this limit are analyzed in Sections~\ref{sec:2drf} and \ref{sec:micromag}. We note that because of the negligible anisotropy of Py, the local magnetic configuration in such a state is determined entirely by the competition between the random field and the exchange stiffness. Therefore, the magnetization in such a state is twisted everywhere, i.e. it is an xy version of the HDS predicted by Malozemoff.

Here, we consider the opposite limit of quasi-uniform $h_\perp$, $l_h>l_M$, such that the first term in Eq.(\ref{eq:small_angles}) can be neglected. This limit may provide a good description for the exchange-spring behaviors of thin-film polycrystalline AFs, where the characteristic length scales for the variation of interfacial exchange torques, determined by the ``winding" of the exchange spring, are expected to be determined by the size of AF grains~\cite{PhysRevB.59.3722,PhysRevLett.92.247201}.

In this limiting case, $\varphi=h_\perp/(H+h_\parallel)$, i.e. $\vec{M}(\vec{r})$ is simply aligned with the local net effective field $\vec{H}+\vec{h}$, as illustrated by the simulations in Figs.~\ref{fig:randcorr}(c),(d). For the average magnitude of deviation from saturation, we obtain

\begin{equation}\label{eq:Mperp}
\left<\varphi^2\right>=\frac{\left<h^2_\perp\right>}{(H+h_\parallel)^2},
\end{equation}
where we have neglected the higher-order effects associated with the spatial variations of $h_\parallel$. This approximation is justified, for example, for $H\gg h_\parallel$.

By fitting the experimentally determined dependence of $\left<\varphi^2\right>$ on $H$ with Eq.(\ref{eq:Mperp}), one can determine the parameters $\left<h^2_\perp\right>$ and $h_\parallel$. In the discussion and figures presented in the next section, we will for brevity use the notation $h_\perp$ when referring to $\sqrt{\left<h_\perp^2\right>}$. For $l_h\gg l_M$, both $h_\parallel$ and $h_\perp$ are expected to scale inversely with the thickness $t$ of the ferromagnet [see Eq.~(\ref{eq:Heff})]. Some of the data discussed below exhibit significant deviations from this expected dependence. We will present analysis based on a combination of analytical calculations, simulations, and scaling, to show that these results are consistent with Malozemoff's hypothesis of uncorrelated random effective exchange field.

\section{Experiment}\label{sec:exp}

Multilayer films with the structure CoO(6)Py($t$)Ta(5) were deposited on $6$~mm$\times 2$~mm silicon substrates at room temperature, in a high vacuum sputtering system with the base pressure of $5\times10^{-9}$~Torr. The numbers in parenthesis are thicknesses in nanometers, the thickness $t$ of Py was varied between $5$~nm and $50$~nm, and Ta(5) served as a capping layer protecting the films from oxidation. The multilayers were deposited in $150$~Oe in-plane magnetic field, which is known to facilitate magnetic ordering in CoO. Py and Ta were deposited by dc sputtering from the stoichiometric targets, in $1.8$~mTorr of ultrapure Ar, while CoO was deposited from a Co target by reactive sputtering in ultrapure oxygen atmosphere, with the partial pressure of oxygen optimized as in our previous studies of CoO-based systems~\cite{PhysRevB.78.052403,PhysRevB.94.024422,urazhdin2019JMMM}.

\begin{figure}
	\includegraphics[width=\columnwidth]{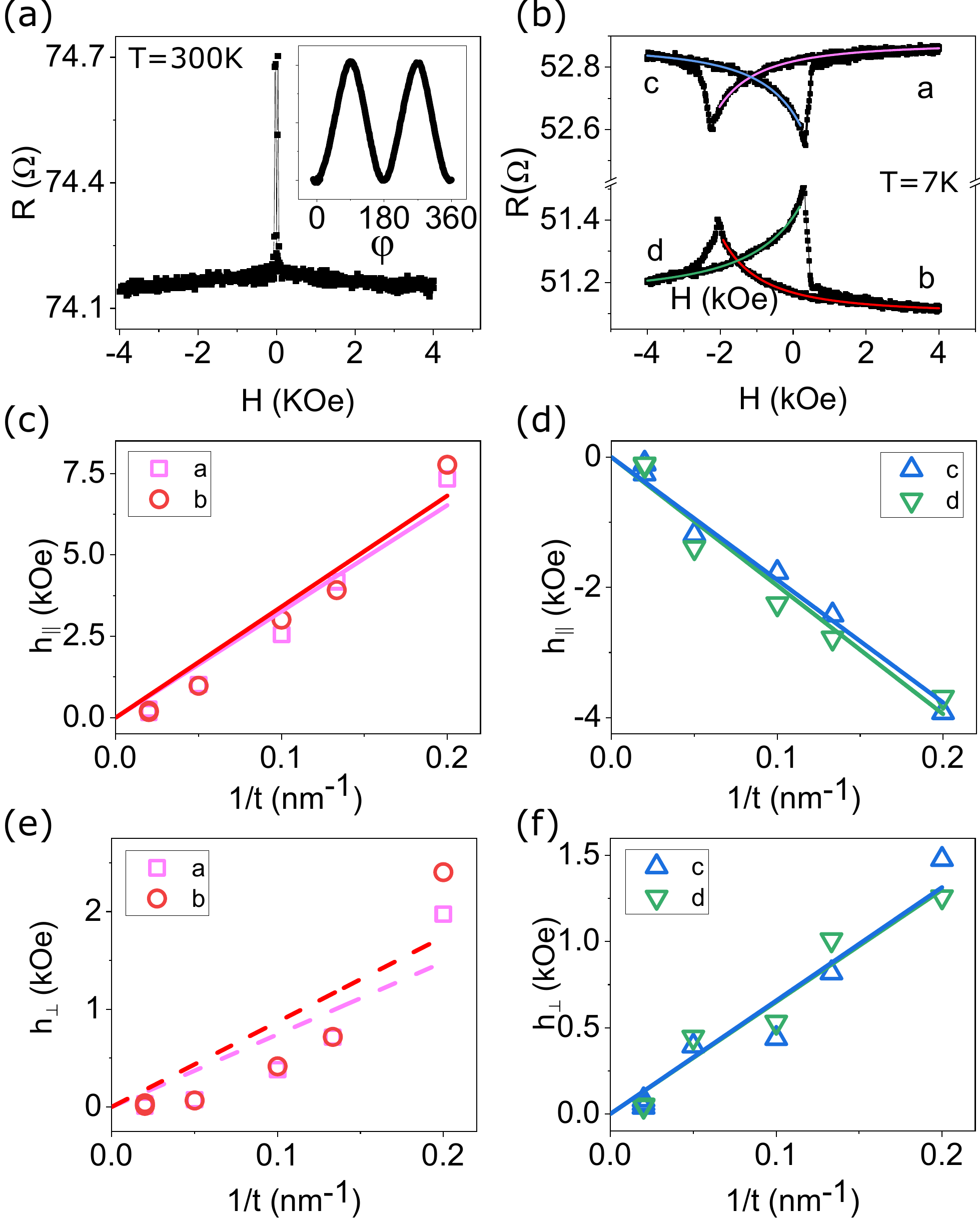}
	\caption{Evidence for random-field effects in Py/CoO bilayers. 
		(a) Magnetoelectronic hysteresis loop of Py(7.5)/CoO(6) measured at $300$~K, with the external field $\vec{H}$ oriented in-plane perpendicular to the current. Inset: dependence of resistance on the direction of in-plane field $H=1$~kOe, at $T=300$~K. (b) Symbols: Magnetoelectronic hysteresis loop for Py(7.5)/CoO(6) at $T=7$~K, for external field parallel to current (labeled a and c) and perpendicular  to current (labeled b andd). Curves: fits with Eq.~(\ref{eq:Mperp}). (c)-(f) Symbols: $h_\parallel$ (c), (d) and $ h_\perp$ (e), (f) vs $1/t$ obtained from the fits as shown in (b), for the four hysteresis branches a-d. Lines are linear fits with zero intercept.}\label{fig:exp1} 
\end{figure}

To characterize the unsaturated magnetization state of the Py films in the studied heterostructures, we utilized electronic measurements of the variations of resistance $R$ due to the anisotropic magnetoresistance (AMR), using ac current with rms amplitude of $50$~$\mu$A and lock-in detection in the four-probe van der Pauw geometry. The AMR exhibits a $180^\circ$-periodic sinusoidal dependence on the angle between the magnetization of Py and the direction of current, as was verified by measurements at temperature $T=300$~K above the Neel temperature of CoO, $T_N=291$~K [inset in Fig.~\ref{fig:exp1}(a)].

Measurements described below were performed for two orientations of the external field, one collinear and the other perpendicular to the direction of current, so that in the saturated state the AMR was maximized and minimized, respectively. Any deviations from saturation resulted in resistance decrease in the first configuration, and increase in the other. These were the signals detected in our magnetoelectronic measurements to characterize the inhomogeneous states. Data analysis was limited only to resistance ranges deviating by less than $10\%$ of the full magnetoresistance from the saturation value, ensuring the small-angle limit for $\varphi$. For the measurements performed at $T<T_N$, the sample was cooled through $T_N$ in field $H=1$~kOe. The cooling field was aligned with the positive direction of the field $H$ utilized in the subsequent measurements.

At high temperature $T>T_N$, CoO is a paramagnet, and is not expected to significantly affect the state of Py. The magnetization $\vec{M}$ of Py is expected to become saturated at small fields determined by the magnetocrystalline anisotropy of Py. Indeed, magnetoelectronic hysteresis loop measurements show negligible variations of $R$, aside from a sharp peak at small $H$ associated with the reversal of $M$, as shown in Fig.~\ref{fig:exp1}(a) for Py(7.5)/CoO(6). In contrast, at $T=7$~K, the $R$ vs $H$ curves exhibit gradual variations and do not saturate even at $H=\pm 4$~kOe, Fig.~\ref{fig:exp1}(b).

These data clearly indicate the presence of a large transverse component $H_\perp$ of the effective exchange field, resulting in the deviations of magnetization from the saturated state even at large $H$. The curves labeled a,c were acquired using the field direction collinear with the current direction, such that the resistance is maximized when $M$ is saturated along the field. Meanwhile, the curves labeled b,d were acquired with the  field perpendicular to the current, resulting in the resistance minimum in the saturated state. These two complementary sets of measurements are necessary for the quantitative data analysis, as discussed below.

The peaks in the hysteresis curves correspond to the magnetization reversal points. These points are shifted in the negative-field direction in Fig.~\ref{fig:exp1}(b), as expected due to the exchange bias effect. We note that the values of $R(H)$ do not exactly coincide for two opposite directions of field sweep. The difference can be attributed to the aging phenomena in AF, as demonstrated by recent time-domain measurements~\cite{PhysRevB.94.024422}. Aging effects were shown to be large for CoO thicknesses below $4$~nm, and become rapidly reduced for larger thicknesses. To minimize their possible influence on our analysis, we focus below only on the hysteresis branches obtained with the field swept from larger to smaller magnitudes.

To directly relate our $R(H)$ data to the analysis presented above, we note that AMR provides direction information about the local deviations of the magnetization state from saturation, according to $R=R_{min}+\Delta Rsin^2\varphi$ for $\vec{H}$ perpendicular to the current, and $R=R_{max}-\Delta Rsin^2\varphi$ for $\vec{H}$ parallel to the current. Here, $R_{min}$ and $R_{max}$ are the minimum and the maximum of resistance due to AMR, respectively, $\Delta R=R_{max}-R_{min}$, and $\varphi(\vec{r})$ is the angle between $\vec{H}$ and $\vec{M}$. For $h_\perp$ characterized by a large correlation length $l_h$, we obtain from Eq.(\ref{eq:Mperp}) for small $\varphi$
\begin{equation}\label{eq:Rpar}
R=R_{max}-\Delta R \frac{h_{\perp}^2}{(H+h_{\parallel})^2},
\end{equation}
for the external field direction parallel to current, and 
\begin{equation}\label{eq:Rperp}
R=R_{min}+\Delta R \frac{h_{\perp}^2}{(H+h_{\parallel})^2},
\end{equation}
for the external field perpendicular to current. We emphasize that Eqs.~(\ref{eq:Rpar}), (\ref{eq:Rperp}) are valid only in the limit of large correlation length $l_h$ of $\vec{h}$, so that the magnetization locally follows the direction of the total effective field.

The curves in Fig.~\ref{fig:exp1}(b) show the results of data fitting with Eqs.(\ref{eq:Rpar}) and (\ref{eq:Rperp}), with $h_\parallel$ and $h_\perp$ treated as independent parameters for each of the four branches, but with the same fitting values of $R_{min}$, $R_{max}$, and $\Delta R=R_{max}-R_{min}$. By fitting all the four branches of the hysteresis loops obtained for different thicknesses $t$ of Py with Eqs.~(\ref{eq:Rpar}) and (\ref{eq:Rperp}), the dependence of  $h_\parallel$  and $h_{\perp}$ on $t$ was determined. Since both of these quantities represent the effects of exchange interaction at the F/AF interface averaged over the thickness of Py, they are expected to scale inversely with $t$ [see Eq.~(\ref{eq:Heff})].  To assess the validity of this expectation, we plot the dependences of $h_{\parallel}$ and $h_{\perp}$ on $1/t$ in Figs.~\ref{fig:exp1}(c),(d) and Figs.~\ref{fig:exp1}(e),(f), respectively.

The dependence $h_\parallel(1/t)$ is well described by a linear fit with zero intercept for all four branches [Figs.~\ref{fig:exp1}(c),(d)], consistent with our analysis. We emphasize that this result is expected regardless of the correlation length $l_h$ of the effective exchange field, because the spatial average of $h_\parallel(\vec{r})$ is finite. Similarly, $h_\perp(1/t)$ is also well described by a linear fit with zero intercept, for the hysteresis branches c,d corresponding to the magnetization state reversed relative to the field-cooling, Fig.~\ref{fig:exp1}(f). This result indicates that the correlation length $l_h$ of the effective exchange field is large in this reversed state, consistent with the picture of AF exchange spring ``wound'' by the reversal of magnetization, with the same ``winding'' direction over a significant volume of CoO the may include the entire grains of the polycrystalline CoO film~\cite{PhysRevB.59.3722,PhysRevLett.92.247201}.  

In contrast, for the two branches a,b corresponding to the magnetization aligned with the field-cooling direction, the dependence $h_{\perp}(1/t)$ is strongly nonlinear [Fig.~\ref{fig:exp1}(e)], demonstrating that the correlated effective exchange field approximation underlying Eqs.(\ref{eq:Rpar}) and (\ref{eq:Rperp}) is invalid. We emphasize that the linear fits in this panel are included only to highlight the nonlinear variations of the data. These fits are not used in this work to determine any physically meaningful parameters of the studied system.

The values of $h_{\perp}(1/t)$ extracted from our analysis increase superlinearly with increasing  $1/t$. This result can be qualitatively expected for the effects of random field with a small correlation length, because at large $1/t$ (small $t$), magnetic correlations within F are less efficient in averaging the short-scale variations of the field. To quantitatively analyze our results, in the next sections we will extend our analysis of the magnetization state of F in F/AF bilayer to include the effects of random uncorrelated effective fields, and show that the results of Fig.~\ref{fig:exp1}(d), for the field parallel to the cooling field, are consistent with the presence of uncorrelated random effective exchange field at the Py/CoO interface.

\section{2d xy model of uncorrelated random field effects}\label{sec:2drf}

In this section, we analyze the effects of an uncorrelated random field on a 2d magnetic system. This analysis is expected to be applicable to magnetic films with sufficiently small thickness $t$, such that their magnetization is uniform through the thickness. In the next section, we present realistic 3d micromagnetic simulations of thin films, and show that their results asymptotically approach our analytical  predictions for 2d systems in the limit of vanishing film thicknesses.

Since Py is characterized by negligible magnetocrystalline anisotropy, and its magnetization in the studied films remains in-plane due to the large demagnetizing field, the system can be described by the 2d xy model. We follow the approaches of Chudnovsky, who analyzed the effects random field on the 2d Heisenberg model~\cite{Chudnovsky1983}, and of Garanin et al., who analyzed the 3d version of a similar random-field xy model~\cite{PhysRevB.88.224418}. The system is characterized by the position-dependent angle $\varphi(\vec{r})$ between the magnetization and the external field, which is determined by the distribution of the effective field $\vec{h}(\vec{r})$ according to Eq.~(\ref{eq:small_angles}). The average of the component $h_\parallel$ of the effective field parallel to $\vec{H}$ , which is nonzero in the experimental system discussed in this paper, is absorbed into the definition of $H$. Thus, in the analysis below, we assume that both $h_\parallel$ and $h_\perp$ form the same random distributions with zero averages. Since $\varphi$ is small at sufficiently large $H$, the term $\varphi h_\parallel$ in Eq.~(\ref{eq:small_angles}) can be neglected, giving

\begin{equation}\label{eq:rf1}
\frac{A}{\mu_0M}\nabla^2\varphi-\varphi H=-h_\perp.
\end{equation}

The random field $h_\perp$ is assumed to be uncorrelated among different lattice sites $i,j$, $\left<h_{\perp,i}h_{\perp,j}\right>=h^2\delta_{ij}/2$. In the micromagnetic simulations discussed in the next section, the simulation cells play the role of the lattice sites. To capture the effects of random field, the cubic cell size $D$ must be smaller than the magnetic correlation length $l_M$. The magnitude of the random field is then scaled between the two descriptions according to $h_{\perp,mm}D=h_{\perp,at}\sqrt{P}$, where $P$ is the area per site of the 2d lattice, $\sqrt{P}=a$ for square lattices, and $\sqrt{P}=a/4\sqrt{3}$ for the (111) face of the fcc lattice. In the continuous limit discussed in this section,  
\begin{equation}\label{eq:corr}
\left<h_\perp(\vec{r})h_\perp(\vec{r}')\right>=h^2P\delta(\vec{r}-\vec{r}')/2.
\end{equation}

Using $k=1/l_M=\sqrt{\mu_0MH/A}$, we rewrite Eq.~(\ref{eq:rf1}) as 
\begin{equation}\label{eq:rf2}
(\nabla^2-k^2)\varphi=-h_\perp\mu_0M/A.
\end{equation}
The solution in terms of the Green's function $G(k,\vec{r})$ of the operator $\nabla^2-k^2$ is
\begin{equation}\label{eq:varphi}
\varphi(\vec{r})=-\frac{\mu_0M}{A}\int d^{2}\vec{r}'G(k,\vec{r}-\vec{r}')h_\perp(\vec{r}').
\end{equation}

The Green's function can be expressed in terms of the modified Bessel function of the second kind, $K_{0}(x)=\frac{1}{2}\int_{-\infty}^{+\infty}\frac{e^{ixt}dt}{\sqrt{1+t^{2}}}$, $G(k,\vec{r})=-K_0(k|r|)/2\pi$. The average of $\varphi^2$ over the realizations of random field is

\begin{equation}\label{eq:varphi_ave}
\begin{split}
\left<\varphi^2(\vec{r})\right>=\left(\frac{\mu_0M}{2\pi A}\right)^2\int d^{2}\vec{r}'d^{2}\vec{r}''K_0(k|\vec{r}-\vec{r}'|)\cdot\\
\cdot K_0(k|\vec{r}-\vec{r''}|)\left<h_\perp(\vec{r}')h_\perp(\vec{r}'')\right>.
\end{split}
\end{equation}

Using the correlation relation Eq.~(\ref{eq:corr}), we obtain
\begin{equation}\label{eq:varphi_ave}
\left<\varphi^2(\vec{r})\right>=\frac{\mu_0^2M^2h^2P}{8\pi^2A^2}\int d^{2}\vec{r}'K^2_0(k|\vec{r}-\vec{r}'|).
\end{equation}
 
Finally, we use the relation $\int d^2rK^2_0(kr)=\pi/k^2$ to obtain
\begin{equation}\label{eq:2dxy_result}
\left<\varphi^2\right>=\frac{\mu_0^2M^2h^2P}{8A^2k^2}=\frac{\mu_0Mh^2P}{8AH}.
\end{equation}

In comparison, Garanin et al.~\cite{PhysRevB.88.224418} obtained $\left<\varphi^2\right>\propto h^2/\sqrt{H}$ for the 3d xy random field model, and our correlated-random-field result, Eq.~(\ref{eq:corr}), is $\left<\varphi^2\right>\propto  h^2/H^2$. In all cases, $\left<\varphi^2\right>\propto h^2$. This can be expected from the general Eq.~(\ref{eq:rf1}) for the magnetization distribution, which is invariant under the scaling transformation $h_\perp\to\alpha h_\perp$, $\varphi\to\alpha\varphi$. Thus, this result is expected to generally hold regardless of the system geometry or the spatial properties of $\vec{h}$. On the other hand, these expressions contain different powers of external field $H$, dependent on the random field distribution and the dimensionality of the system. All these relations can be written in an explicitly dimensionless form as 

\begin{equation}\label{eq:scaling}
\left<\varphi^2\right>=C\left(\frac{h}{H}\right)^2\left(\frac{P}{l_M^2}\right)^d,
\end{equation}
where the numeric coefficient $C$ and the power-law exponent $d$ are dependent on the system realization. For the correlated random field, $d=0$, while for the uncorrelated random field in 2d (3d), $d=1$ ($3/2$). Based on the scaling arguments for the random field, we expect $d=n/2$ for the uncorrelated random field in $n$ dimensions. In the next section, we use Eq.~(\ref{eq:scaling}) as an ansatz with $d$ treated as a fitting parameter, to analyze the micromagnetic simulations of interfacial exchange effects in F/AF bilayers.

\section{Simulations of uncorrelated random field effects}\label{sec:micromag}

The analytical model introduced in the previous section is expected to quantitatively describe the effects of uncorrelated random field only for atomically-thin F. For finite thickness $t$ of F in F/AF bilayers, magnetic moments away from the F/AF interface experience only indirect effects of effective exchange field averaged over their neighbors, introducing spatial correlations that are not accounted for by the model. In this section, we use 3d micromagnetic simulations and an extension of the scaling arguments presented above to analyze a more realistic model where random field is applied only to one of the surfaces of a thin Py film. We also show that the results are consistent with the analytical model in the limit of ultrathin films. 

We performed micromagnetic simulations with the mumax3 software~\cite{vansteenkiste2014design}, using the standard parameters for Py, the magnetization $\mu_0M=1.0$~T, Gilbert damping $\alpha=0.01$, and exchange stiffness $A=1.3\times10^{-11}$~J/m. The simulated volume was $ 2~\mu m \times 2~\mu m\times t$, with varied thickness $t$. This volume was discretized into cubic cells, whose size $D$ was varied from $1$~nm to $12$~nm to evaluate the discretization effects, as described below. Periodic boundary conditions were used to eliminate edge effects. Random uncorrelated field with fixed magnitude $h$ was generated by selecting a random variable $\phi$ uniformly distributed over the interval $[0,2\pi]$. In all the simulations discussed below, this field was applied only to the bottom layer of the simulation mesh.

In the limit of vanishing film thickness, $D\to0$ and only one layer present in the simulation mesh, this system maps onto the analytical model described in the previous section via $D^2=P$. The magnitude of $h$ can be related to the effective exchange field experienced by the atoms at the interface, according to $H_{int}=3^{3/4}2hD^2/a^2$ for the (111)-textured surface of fcc ferromagnet with a cubic lattice constant $a$.

\begin{figure}
	\includegraphics[width=\columnwidth]{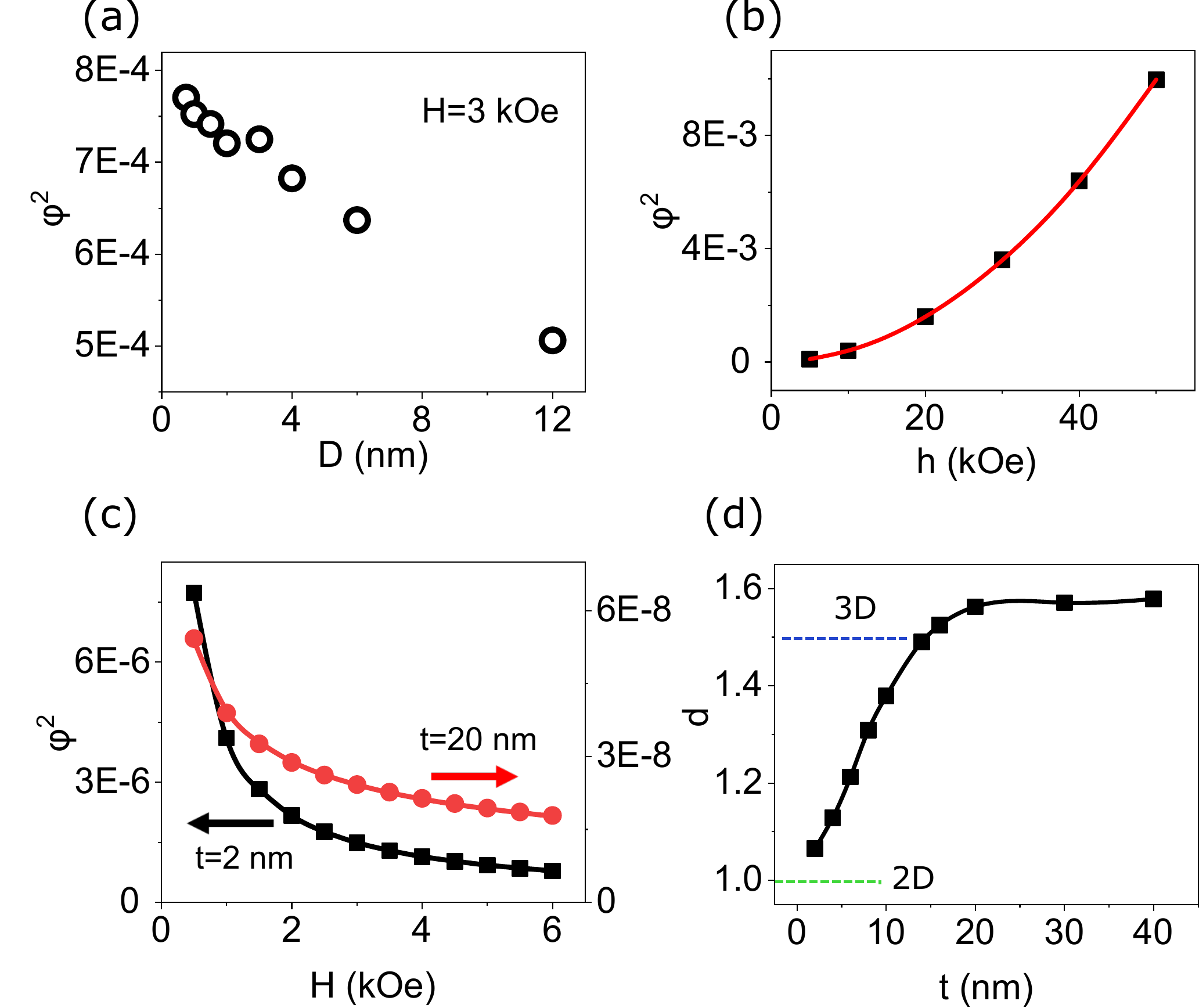}
	\caption{Micromagnetic simulations of random field effects.
		(a) $\left<\varphi^2\right>$ vs cell size $D$ for a $12$~nm-thick Py film, at $H=3$~kOe and $\mu_0hD^2=5~T\cdot nm^{2}$. (b) Symbols: $\left<\varphi^2\right>$  vs $h$ for a $10$~nm-thick Py film, at $H=6$~kOe. Curve: fit with a quadratic function. (c)  $\left<\varphi^2\right>$ vs $H$, for Py films with $t=2$~nm and $t=20$~nm, as labeled. Symbols are the results of simulations, and curves are fits using the ansatz Eq.~(\ref{eq:scaling}). (d) Dependence of the power law exponent $d$ in Eq.~(\ref{eq:scaling}) on the Py film thickness.
	}\label{fig:simulations} 
\end{figure}

The simulations were performed with the magnetic system initialized in a uniform state aligned with the field $\vec{H}$, and were continued until the dynamics became negligible for all the simulation cells. The distribution was then analyzed to determine $\left<\varphi^2\right>$. Figs.~\ref{fig:randcorr}(a),(b) illustrate a representative random field distribution and the resulting magnetization map in the equilibrium state, for $t=D=2$~nm, $H=4$~kOe, $h=50$~kOe. While the random field distribution is uncorrelated, the resulting magnetization distribution exhibits correlations on the length scale $l_M=\sqrt{A/\mu_0MH}=6$~nm. For the correlated field with the correlation length $l_h>l_M$, the magnetization is expected to simply follow the local direction of the net effective field, as was verified by the simulation using random field with correlation length $l_h=18$~nm [Figs.~\ref{fig:randcorr}(c),(d)].

To determine the optimal simulation cell size $D$ that does not significantly distort the magnetization response to the random field, we performed simulations with different values of $D$ ranging from $1$~nm to $12$~nm, Fig.~\ref{fig:simulations}(a). To facilitate direct comparison, the value of $h$ was adjusted so that $hD^2$ remained independent of $D$, in accordance with the scaling relations expected for the random field. The value of $\left<\varphi^2\right>$ monotonically decreases with increasing $D$, as expected due to the filtering effect of larger cells on the short-scale random field variations. In the simulations discussed below, we use a sufficiently small cell size $D=2$~nm so that these filtering effects are small, while keeping the simulations of thicker films manageable.

Figure~\ref{fig:simulations}(b) shows the dependence of $\left<\varphi^2\right>$ on $h$, with all the other parameters fixed. This dependence is precisely described by the quadratic relation expected from Eq.~(\ref{eq:scaling}). Thus, it is sufficient to perform simulations only for one value of $h$ small enough to satisfy the weak random field approximation $\varphi^2\ll 1$. 

The central goal of our simulations was to determine the dependence of random field effects on the film thickness. To this end, we performed simulations of the dependence of the magnetization state on the external bias field $H=0.5-6$~kOe for thicknesses $t=2-40$~nm, with $h$ fixed at $100$~Oe. In all cases, the dependence of $\left<\varphi^2\right>$ on $H$ could be precisely fitted by Eq.~(\ref{eq:scaling}), or equivalently 
\begin{equation}\label{eq:mum_fit}
\left<\varphi^2\right>=C'\frac{h^2D^4}{H^{2-d}},
\end{equation}
with the power-law exponent $d$ and the constant $C'=CD^{-4}(\mu_0Ma^2/4\sqrt{3}A)^d$ used as fitting parameters. In this expression, we scaled $h$ by the cell size, so that the constant $C'$ becomes independent of $D$. Figure~\ref{fig:simulations}(c) shows the fits for two representative thicknesses $t=2$~nm and $20$~nm, yielding the best-fit values $d=1.065$ and $1.57$, respectively. We note that these two representative dependences are substantially different, demonstrating that precise fitting requires the value of $d$ to be varied with $t$.

Figure~\ref{fig:simulations}(d) shows the dependence of the power-law exponent $d$ on the film thickness, extracted from the $\left<\varphi^2\right>$ vs $H$ curves such as those shown in Fig.~\ref{fig:simulations}(c). This dependence extrapolates to $d=1$ in the limit of vanishing film thickness, consistent with the results of the analytical 2d xy model described in the previous section. The value of $d$ increases with $t$, reaching $d_s=1.57$ for $t=20$~nm, and becomes constant at larger $t$. Qualitatively, these behaviors can be interpreted in terms of the crossover from the effective 2d regime to the effective ``bulk" regime, where the effects of random field become almost completely averaged out far enough from the interface, such that increasing $t$ simply rescales $\left<\varphi^2(H)\right>$ due to averaging over the larger volume, without changing the functional relation. We emphasize that random field is applied only to one of the film surfaces. Thus, this regime is not equivalent to the 3d random-field model considered by Garanin et al.~\cite{PhysRevB.88.224418}. Indeed, the saturation value $d_s$ is different from $d=3/2$ obtained in the latter case.

\section{Analysis of experimental results}\label{sec:analysis}

We now show that Eq.~(\ref{eq:scaling}), with the power-law exponent $d(t)$ determined from the micromagnetic simulations, provides an explanation of our experimental data, supporting Malozemoff's uncorrelated random-field hypothesis.

If the effects of the exchange field at the Py/CoO interface can be approximated by a random field uncorrelated on the atomic scale, then the dependence of $R$ on $H$ can be inferred from Eq.~(\ref{eq:mum_fit}), with the power-law exponent $d$ and the scaling constant $C'$ determined from the simulations discussed above, $H$ offset by $h_\parallel$, and $h^2D^4$ replaced by $H_{int}^2a^4/4\sqrt{3}$,
\begin{equation}\label{eq:Rpar2}
R=R_{max}-\frac{C'\Delta R}{4\sqrt{3}}\frac{H_{int}^2a^4}{(H+h_\parallel)^{2-d}},
\end{equation}
for the external field parallel to current, and 
\begin{equation}\label{eq:Rperp2}
R=R_{min}+\frac{C'\Delta R}{4\sqrt{3}}\frac{H_{int}^2a^4}{(H+h_\parallel)^{2-d}},
\end{equation}
for the external field perpendicular to current.

\begin{figure}
	\includegraphics[width=\columnwidth]{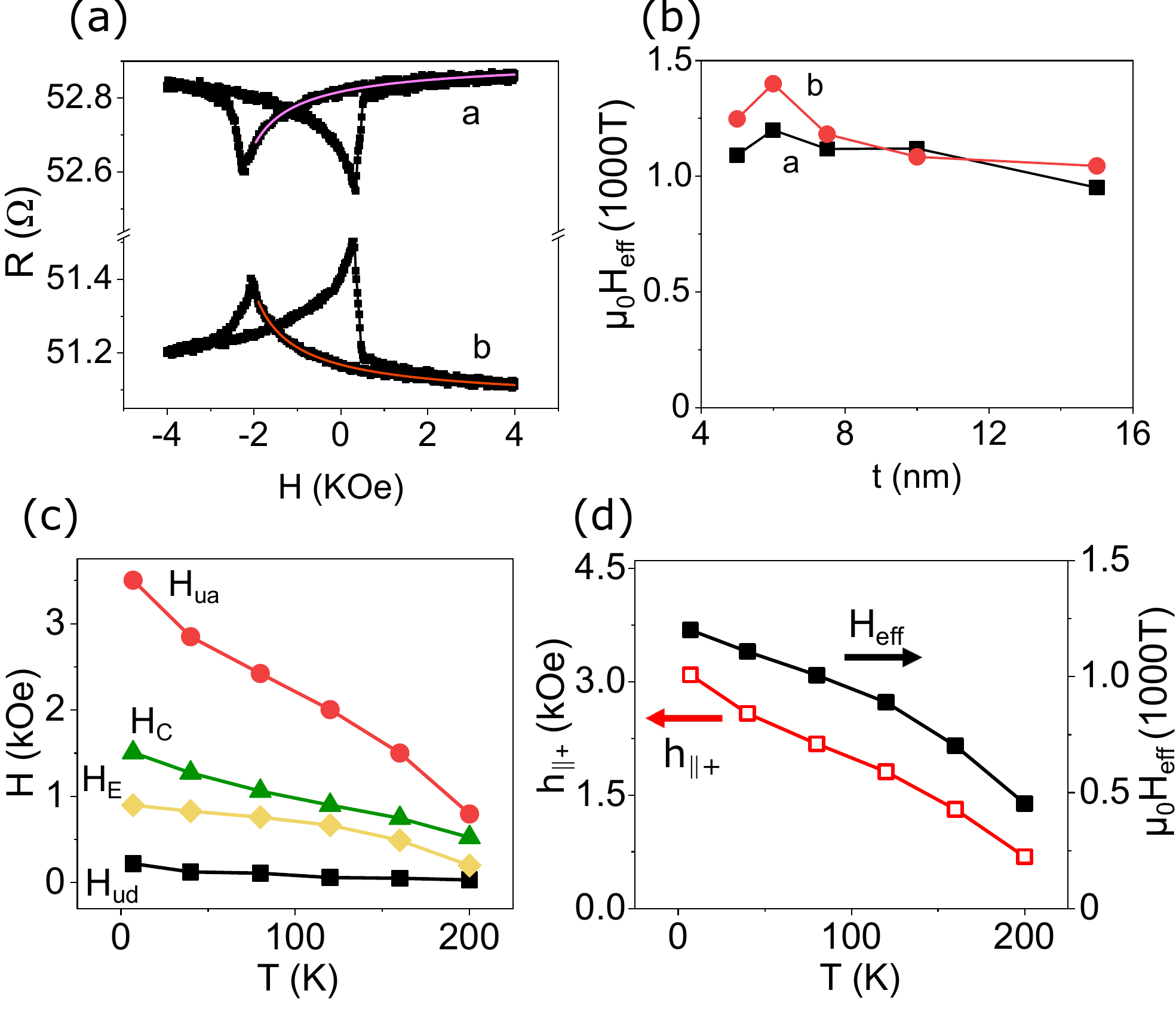}
	\caption{Quantitative analysis of effective exchange fields. (a) Symbols: the same magnetoelectronic hysteresis loop as in Fig.~\ref{fig:exp1}(b), acquired at $T=7$~K for Py(7.5)/CoO(6). Curves: fits of branches a,b based on Eq.~(\ref{eq:scaling}), with the power-law exponent $d=1.28$ determined from the micromagnetic simulations. (b) The magnitude of the effective random exchange field $\mu_0H_{eff}$ vs Py thickness, determined from fits such as shown in panel (a). (c) Coercivity $H_C$, effective exchange bias field $H_E$, effective uniaxial anisotropy field $H_{ua}$, and unidirectional anisotropy field $H_{ud}$ vs $T$, determined for Py(6)/CoO(6) as discussed in the text. (d) Parallel component $h_{\parallel,+}$ of the effective exchange field, [open symbols and right scale] and the effective random field $H_{eff}$ [solid symbols and right scale] vs $T$ for Py(6)/CoO(6), obtained from branch $a$ of the $R$ vs $H$ data.}\label{fig:analysis}
\end{figure}

Figure~\ref{fig:analysis}(a) shows the same data as in Fig.~\ref{fig:exp1}(b), but now fitted using Eqs.~(\ref{eq:Rpar2}), (\ref{eq:Rperp2}), with the power-law exponent $d=1.28$ for Py(7.5) determined from the micromagnetic simulations described above. Both this fitting and the fitting with $d=0$ in Fig.~\ref{fig:exp1}(b) provide good fits for the data. This shows that, in contrast to the micromagnetic simulations, the power-law exponent $d$ cannot be accurately determined from the experimental data. The reason for this discrepancy is that the values $R_{min}$ and $R_{max}$ of resistance in the saturated states with the magnetization perpendicular and parallel to current, respectively, as well as the parallel component $h_\parallel$ of the effective exchange field, cannot be independently determined, and must be thus treated as additional fitting parameters. The experimental data do not provide sufficient information to accurately determine these parameters together with $d$. 

While fitting the experimental $R$ vs $H$ curves does not allow us to determine $d$, we can still establish whether the observed behaviors are consistent with the uncorrelated random field approximation. We use the approach similar to that described in Section~\ref{sec:exp}, where we have shown that the correlated effective field approximation cannot describe the magnetization state for the field aligned with the cooling field [see Fig.~\ref{fig:exp1}(d)]. We fit the $R(H)$ curves for different thicknesses $t$ of Py with Eqs.~(\ref{eq:Rpar2}), (\ref{eq:Rperp2}), using the thickness-dependent values of $d(t)$ and $C'(t)$ obtained from the micromagnetic simulations. Each such fitting independently yields the value of the effective exchange field $H_{int}$. The uncrorrelated random field approximation is valid if the obtained values of $H_{int}$ are independent of $t$. However, if the effective exchange field is correlated, then the values of $H_{int}$ extracted from such fitting should increase with $t$, because in contrast to the uncorrelated field, the effects of the correlated field are not averaged out by larger thickness. 

Figure~\ref{fig:analysis}(b) shows the values of $\mu_0H_{int}$ determined from the fits of $R(H)$ for different Py thicknesses. The values exhibit modest variations around the average value of $1\times10^3$~T, and appear to slightly decrease at large $t$, but clearly do not increase, as would be expected for the correlated field. We note that our procedure for calculating the values of $H_{eff}$ involves multiple sources of random and systematic errors, including the uncertainty of the thicknesses of Py, slight variations of the deposition conditions resulting in the variation of $H_{eff}$ among different samples, as well as the uncertainty of the fitting itself. These uncertainties are difficult to estimate {\it a priori}, warranting more detailed studies of multiple similar samples to assess them statistically. Nevertheless, the results shown in Fig.~\ref{fig:analysis}(b) for five samples with different thicknesses provide strong evidence for the validity of random-field approximation.
Furthermore, the magnitude of $\mu_0H_{eff}$ of about $1\times10^3$~T is about 10 times smaller than the typical strength of the nearest-neighbor exchange interactions in magnetic materials~\cite{OHandley}, as would be expected given that the spin-flop of AF spins at the F/AF interface results in their partial alignment~\cite{Koon97,Schulthess98}. 

Our approach to quantifying the effective exchange fields in F/AF bilayers is validated by the analysis of the relationship between these fields and the essential characteristics of the magnetic hysteresis  loop, the coercivity $H_C=(H_1-H_2)/2$ and the exchange bias field $H_E=(H_1+H_2)/2$. Here, $H_1$ ($H_2$) is the magnetization reversal field on the down (up) sweep, signified by the sharp peaks in $R$ vs $H$ curves [see Fig.~\ref{fig:analysis}(b)]. The exchange bias field is generally attributed to the unidirectional anisotropy, while the enhanced coercivity is attributed to the uniaxial anisotropy acquired by F due to the exchange interaction at the F/AF interface. 

Our approach allowed us to determine the value of $h_\parallel$, the net effective exchange field experienced by Py, separately for the magnetization orientation parallel to the cooling field [by fitting $R(H)$ branches a,b with Eqs. (\ref{eq:Rpar2}), (\ref{eq:Rperp2})], and 
for the magnetization orientation opposite to the cooling field [by fitting $R(H)$ branches c,d with Eqs. (\ref{eq:Rpar}), (\ref{eq:Rperp})]. We label the corresponding two values $h_{\parallel,+}$ and $h_{\parallel,-}$. The effective unidirectional and uniaxial anisotropy fields can be then directly determined as $H_{ud}=(h_{\parallel,+}+h_{\parallel,-})/2$ and $H_{ua}=(h_{\parallel,+}-h_{\parallel,-})/2$, respectively. We emphasize that these values are determined by fitting the $R(H)$ curves for small deviations from saturation at large fields, completely independently from $H_C$, $H_E$ that characterize magnetization reversal at small fields.

Figure~\ref{fig:analysis}(c) shows the temperature dependences of all four characteristics $H_E$, $H_C$, $H_{ud}$, and $H_{ua}$, for the Py(6)/CoO(6) sample at $T\le 200$~K. At higher temperatures, the deviations from saturation were too small to reliably determine $h_\parallel$ by fitting the $R(H)$ curve. The relations among $H_E$, $H_C$, $H_{ud}$, and $H_{ua}$ are consistent with the results for a similar Py/CoO bilayer system, obtained by a completely different technique of transverse ac susceptibility~\cite{urazhdin2019JMMM}. In particular, that study showed that the unidirectional anisotropy in this system is much smaller than the effective exchange bias field, and does not follow the temperature dependence of the latter. The data in Fig.~\ref{fig:analysis}(c) are consistent with this observation. 
Transverse ac susceptibility measurements also showed that $H_E$ and $H_C$ are about half of $H_{ua}$, and approximately follow the temperature dependence of the latter. These observations are also confirmed by the results in Fig.~\ref{fig:analysis}(c). While these results may seem surprising, they are consistent with the analysis of Ref.~\cite{urazhdin2019JMMM}, which suggested that the asymmetry of the hysteresis loop for the Py/CoO bilayers is predominantly caused not by the unidirectional anisotropy, but rather by the different mechanisms of magnetization reversal between the two opposite magnetization states stabilized by the uniaxial anisotropy. We discuss the underlying mechanism in Section~\ref{sec:summary}.

The random field $H_{eff}$, determined by fitting branches $a$ and $b$ of the $R(H)$ curve with Eqs.~(\ref{eq:Rpar2}) and (\ref{eq:Rperp2}), decreases with increasing temperature [solid symbols and right scale in Fig.~\ref{fig:analysis}(d)], following the same overall trends as $h_{\parallel,+}$ [open symbols and left scale in Fig.~\ref{fig:analysis}(d)]. The similarity between the behaviors of these two quantities is a manifestation of their common origin from the exchange interaction at the Py/CoO interface.

\section{Summary and conclusions}\label{sec:summary}

To summarize our findings, we have developed a new method for studying random effective exchange fields at magnetic interfaces, which extends the previously developed approaches to analyzing the effects of bulk random effective fields on 2d and 3d systems~\cite{Chudnovsky1983,Tejada1991,PhysRevB.88.224418}.  Our method utilizes measurements of deviations from saturation characterized by $\left<\varphi^2\right>$ - the average of the square of the angle between the magnetization and the external field - which follows a power-law dependence on the applied field with the exponent dependent on the characteristics of the exchange field. For the random effective exchange field correlated on the length scales exceeding the magnetic correlation length, the exponent is different from that for the uncorrelated random field, allowing one to distinguish between these two limiting cases. Moreover, the power-law exponent varies as a function of the film thickness, due to the correlations associated with averaging of the effective random field through the magnetic film thickness. By extension, we expect that the specific value of the power-law exponent for a given film thickness, if known precisely, can be utilized to determine the correlation length of random field. We leave analysis of this possibility to future studies.

We have employed our method to study effective exchange fields at the interfaces of Permalloy/CoO bilayers, a classic ferromagnet/antiferromagnet (F/AF) bilayer system extensively studied in the context of exchange bias. We utilizied magnetoelectronic measurements, in which resistance variations provide direct information about deviations from the saturated magnetization state. Analysis of our measurements required that several additional unknown parameters are determined from the data fitting, which did not allow us to directly determine the power-law exponent characterizing the correlations of random effective exchange field. Nevertheless, using the fact that the strength of the 
interaction at the interface must be independent of the film thickness, we showed that the results for the applied field parallel to the cooling field cannot be explained in terms of a correlated random effective field, but are consistent with the uncorrelated field approximation. In contrast, the results for the magnetic field direction antiparallel to the cooling field were in a reasonable agreement with the correlated effective exchange field approximation.

Qualitatively, we attribute the surprising difference between the characteristics for the two opposite field directions to the exchange-spring effects in CoO, which may produce quasi-uniform exchange torques over length scales comparable to the grain sizes of polycrystalline CoO. We also note that our surprising observations are consistent with a recent observation, for similar Permalloy/CoO bilayers, of qualitatively different reversal mechanisms between the two opposite directions of Py magnetization~\cite{urazhdin2019JMMM}. Specifically, transverse ac susceptibility measurements showed that magnetization reversal from the magnetization direction opposite to the field-cooling direction into the direction aligned with the latter, occurs as soon as its energy becomes higher. Because of the large magnetic anisotropy barrier, such reversal must occur via inhomogeneus intermediate magnetization states, for example by domain wall motion.

On the other hand, reversal from the field-cooling direction was shown to occur only when the anisotropy barrier was almost compensated by the external field, indicating that the domain wall propagation is suppressed in this state, and reversal proceeds via quasi-uniform rotation. Our results complement this picture, providing additional clues about the underlying mechanisms. Indeed, uncorrelated random effective field is expected to result in efficient domain wall pinning, suppressing domain wall propagation. On the other hand, correlated random field, inferred from the analysis for the reversed magnetization state and attributed to the formation of AF exchange spring, may be expected to facilitate reversal through inhomogeneous magnetization state, consistent with the prior observations.

We now discuss the broader impact of our results on the studies and applications of thin magnetic film  systems. First, the effective exchange field in F/AF bilayers, which is the focus of our study, is just one specific case of many magnetic interfacial effects extensively researched and commonly utilized in the existing and emerging technologies. Those include the Ruderman–Kittel–Kasuya–Yosida (RKKY) interaction commonly employed in magnetic multilayer sensors and in artificial antiferromagnets, interfacial magnetic anisotropies commonly utilized to induce perpendicular magnetic anisotropy in magnetic heterostructures, and the interfacial Dzyaloshinski-Moriya interaction~\cite{heck2013magnetic,chappert2010emergence,OHandley}. Understanding the spatial characteristics of these effects is crucial for the development of efficient and reproducible nanodevices. We note that the magnetic anisotropy is equivalent to effective fields for small-angle variations of magnetization, and therefore can be analyzed using the same approach as introduced above. 

Our method becomes particularly effective if the saturation magnetization $M$ of the studied magnetic films is known, and if measurements of deviations from saturation utilize magnetometry, instead of the less direct magnetic characterization by magnetoelectronic techniques used in our study. For almost saturated states, magnetometry provides the value of $(1-\left<\varphi^2\right>)M$, which allows one to directly extract $\left<\varphi^2\right>$, without any additional fitting parameters that were required in our magnetoelectronic measurements. This makes it possible to determine the power-law exponent characterizing the magnetic hysteresis curves, and thus the correlation length of the effective exchange fields, for a single magnetic heterostructure with a specific thickness of the magnetic layer.

Finally, we mention some of the projected fundamental insights that can become facilitated by our work. Our demonstration of uncorrelated effective random field effects in F/AF heterostructures opens the possibility to explore important fundamental consequences of these effects, such as topologically nontrivial magnetization states~\cite{PhysRevLett.112.097201,PhysRevB.88.224418}. Such states can profoundly affect the magnetic properties, but to the best of our knowledge, their effects  in F/AF heterostructures have not yet been explored. Another potentially profound consequence of magnetic frustration associated with uncorrelated effective random fields is the possibility to engineer magnetic energy landscapes whose energy scale is determined by the exchange interaction, rather than the magnetic anisotropy as in unfrustrated magnetic systems. The former is three to four orders of magnitude larger than the latter, providing a unique opportunity to develop ultrasmall thermally stable nanomagnetic devices.

The contributions to this work by G.C. and D.C. were supported by the NSF grant No. ECCS-1804198, the contribution by S.U. was supported by the U.S. Department of Energy (DOE), Basic Energy Sciences (BES), under Award \# DE-SC0018976.
\bibliography{bibl_RF}{}
\bibliographystyle{apsrev4-1}

\end{document}